\documentclass[twocolumn,showpacs,amsmath,amssymb,pre,aps]{revtex4}   
\usepackage{graphicx}
\usepackage{dcolumn}
\usepackage{bm}

\begin{document}
\title{
Critical Currents in Quasiperiodic Pinning Arrays: \\
Chains and Penrose Lattices
}
\draft

\author{Vyacheslav Misko$^{1, 2}$, Sergey Savel'ev$^1$, and Franco Nori$^{1,2}$}
\affiliation{$^1$ Frontier Research System, Institute of Physical and Chemical
Research (RIKEN), Wako-shi, Saitama, 351-0198, Japan}
\affiliation{$^2$ MCTP, CSCS, Physics Department,
University of Michigan, Ann Arbor, MI 48109-1040, USA}

\date{\today}

\begin{abstract}
We study the critical depinning current $J_{c}$ versus the applied
magnetic flux $\Phi$, for quasiperiodic (QP) chains 
and 2D arrays of pinning centers placed on the nodes of a five-fold
Penrose lattice.
In QP chains, 
the peaks in $J_{c}(\Phi)$ are determined by a sequence of harmonics
of the long and short segments of the chain.
The critical current $J_{c}(\Phi)$ has a remarkable self-similarity.
In 2D QP pinning arrays,
we predict analytically and numerically the
main features of
$J_{c}(\Phi)$,
and demonstrate that the Penrose lattice of pinning sites provides
an enormous enhancement of
$J_{c}(\Phi)$,
even compared to triangular and random pinning site arrays.
This huge increase in
$J_{c}(\Phi)$
could be useful for applications.
\end{abstract}
 \pacs{
74.25.Qt 
}
\maketitle

Recent progress in the fabrication of nanostructures
has provided a wide variety of
well-controlled vortex-confinement topologies,
including different regular pinning arrays.
A main fundamental question
in this field was how to drastically increase vortex pinning,
and thus the critical current
$J_{c}$,
using artificially-produced periodic Arrays of Pinning Sites (APS)
\cite{vvmdotprl,fnsc2003,rwdot,vvmbdot,vvmfddot}.
The
increase and, more generally, control
of the critical current
$J_{c}$
in superconductors
by its patterning (perforation)
can be of practical importance for applications
in microelectronic devices.

A peak in the critical current
$J_{c}(\Phi)$,
for a given value of the magnetifc flux per unit cell,
say
$\Phi_{1}$,
can be engineered using a superconducting sample
with a periodic APS
with a matching field
$H_{1}=\Phi_{1}/A$
(where $A$ is the area of the pinning cell),
corresponding to one trapped vortex per pinning site.
However, this peak in
$J_{c}(\Phi)$,
while useful to obtain,
{\it decreases very quickly}
for fluxes away from
$\Phi_{1}$.
Thus, the desired peak in
$J_{c}(\Phi)$
is
{\it too narrow}
and not very robust against changes in
$\Phi$.
It would be greatly desirable
to have samples with APS with {\it many} periods (ideally infinite).
This multiple-period APS sample would provide
either very many peaks or an extremely broad peak in $J_{c}(\Phi)$,
as opposed to just one (narrow) main peak (and its harmonics).
We achieve this goal [a very broad $J_{c}(\Phi)$] here by studying
samples with many built-in periods.

Here,
we study vortex pinning by quasiperiodic (QP) chains 
and by 2D APS located on the nodes of QP lattices
(e.g., a five-fold Penrose lattice) \cite{bookqc}. 
We show that the use of the 2D QP (Penrose) lattice of pinning sites results
in a remarkable
{\it enhancement } of
$J_{c}(\Phi)$,
as compared to other APS, including triangular and random APS.
In contrast to superconducting networks,
for which {\it only} the areas of the network plaquettes play a role
\cite{behrooz},
for vortex pinning by QP pinning arrays,
the specific geometry of the elements which form the QP lattice
{\it and} their arrangement
(and not just the areas)
are important,
making the problem far more complicated.

{\it Simulation.---}
We model a three-dimensional (3D) slab, infinitely long in the $z$-direction,
by a 2D (in the $xy$-plane) simulation cell with periodic
boundary conditions.
We perform simulated annealing simulations by numerically integrating
the overdamped equations of motion (see, e.g.,
Ref.~\cite{md0157,md03Z}):
$
\eta {\rm \bf v}_{i} \ = \ {\rm \bf f}_{i} \ = \ {\rm \bf f}_{i}^{vv} + {\rm \bf f}_{i}^{vp} + {\rm \bf f}_{i}^{T} + {\rm \bf f}_{i}^{d}.
$
Here
${\rm \bf f}_{i}$
is the total force per unit length acting on vortex
$i$,
${\rm \bf f}_{i}^{vv}$
and
${\rm \bf f}_{i}^{vp}$
are the forces due to vortex-vortex and vortex-pin interactions, respectively,
${\rm \bf f}_{i}^{T}$
is the thermal stochastic force,
and
${\rm \bf f}_{i}^{d}$
is the driving force;
$\eta$ is the viscosity, which is set to unity.
The force due to the vortex-vortex interaction is
$
{\rm \bf f}_{i}^{vv} \ = \ \sum_{j}^{N_{v}} \ f_{0} \ K_{1} \!
\left( \mid {\rm \bf r}_{i} - {\rm \bf r}_{j} \mid / \lambda \right)
\hat{\rm \bf r}_{ij},
$
where
$N_{v}$
is the number of vortices,
$K_{1}$
is a modified Bessel function,
$\lambda$
is the penetration depth,
$\hat{\rm \bf r}_{ij} = ( {\rm \bf r}_{i} - {\rm \bf r}_{j} )
/ \mid {\rm \bf r}_{i} - {\rm \bf r}_{j} \mid,$
and
$
f_{0} = \Phi_{0}^{2} / 8 \pi^{2} \lambda^{3} .
$
Here $\Phi_{0} = hc/2e$. 
The pinning force is
$
{\rm \bf f}_{i}^{vp} = \sum_{k}^{N_{p}}  f_{p} \cdot \left( \mid {\rm \bf r}_{i} - {\rm \bf r}_{k}^{(p)} \mid / r_{p} \right) 
\Theta \!
\left[ \left(
r_{p} - \mid {\rm \bf r}_{i} - {\rm \bf r}_{k}^{(p)} \mid \right)/\lambda
\right]
\hat{\rm \bf r}_{ik}^{(p)},
$
where
$N_{p}$
is the number of pinning sites,
$f_{p}$ (expressed in $f_{0}$)
is the maximum pinning force of each short-range parabolic potential well 
located at ${\rm \bf r}_{k}^{(p)}$, 
$r_{p}$
is the range of the pinning potential,
$\Theta$
is the Heaviside step function,
and
$\hat{\rm \bf r}_{ik}^{(p)} = ( {\rm \bf r}_{i} - {\rm \bf r}_{k}^{(p)} )
/ \mid {\rm \bf r}_{i} - {\rm \bf r}_{k}^{(p)} \mid.$
All the lengths (fields) are expressed in units of
$\lambda$ ($\Phi_{0}/\lambda^{2}$).
The ground state of a system of moving vortices is obtained by
simulating the field-cooled experiments (e.g., \cite{tonomura-vvm}).
For deep short-range ($\delta$-like) potential wells,
the energy required to depin vortices 
is proportional to the number of pinned vortices,
$N_{v}^{(p)}$.
Therefore, in this approximation, 
we can define the critical current as follows:
$
j_{c}(\Phi) = j_{0} N_{v}^{(p)}(\Phi) / N_{v}(\Phi),
$
where
$j_{0}$
is a constant,
and study the dimensionless value
$J_{c} = j_{c}/j_{0}$.
We use narrow potential wells as pinning sites,
with
$r_{p} = 0.04\lambda$ to $0.1\lambda$.
We have also performed dynamical simulations of $J_{c}$ using a threshold criterion, 
i.e., $J_{c}$ is obtained as the minimum current $J \  \propto \; f_{\rm i}^{d}$ 
which depins the vortices. 
The results obtained using these two criteria are essentially equal \cite{wepenprb}, 
and here we use the ``static'' criterion defined above.


{\it 1D quasicrystal.---}
A QP chain 
\cite{bookqc} 
can be constructed by
iteratively applying the Fibonacci rule
($
L \rightarrow LS, \ \ \ S \rightarrow L
$),
which generates
an infinite sequence
of
two line segments, long $L$ and short $S$.
For an infinite QP sequence \cite{bookqc}, 
the ratio of the numbers of long to short segments is 
the golden mean 
$
\tau = (1 + \sqrt{5})/2. 
$
Let, e.g., $a_{S}=1$ and $a_{L}=\tau$, where 
$a_{S}$ and $a_{L}$ are the lengths of the short and long segments, respectively. 
Then the position of the $n$th point where a new segment, 
either $L$ or $S$, begins is determined by: 
$
x_{n} = n +  \left[ n/\tau \right] / \tau, 
$
where $\left[ x \right]$ denotes the integer part of $x.$ 
To study the critical depinning current $J_{c}$ in QP
pinning chains, we place pinning sites to the points where the 
$L$ or $S$ elements of the QP sequence link to each other.

\begin{figure}[btp]
\begin{center}
\vspace*{-0.5cm}
\hspace*{-0.5cm}
\includegraphics*[width=9.5cm]{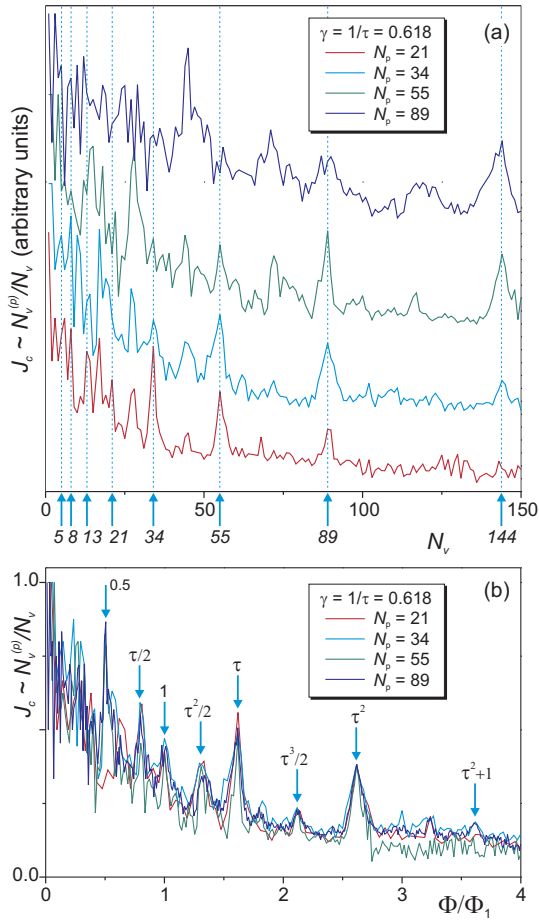}
\end{center}
\vspace*{-1.0cm}
\caption{
(Color online)
(a) The critical depinning current
$J_{c}$,
versus 
the number of vortices,
$N_{v} \sim \Phi$,
for 1D QP chains,
$N_{p} = 21$ (red bottom line),
$N_{p} = 34$ (second from the bottom blue line),
$N_{p} = 55$ (green line),
and
$N_{p} = 89$ (dark blue top line),
for
$\gamma = a_{S}/a_{L} = 1 / \tau$.
Here we use:
$f_{p}/f_{0} = 1.0$ and $r_{p} = 0.1\lambda$.
Independently of the length of the chain,
the peaks
include
the sequence of successive Fibonacci numbers
and their sub-harmonics.
(b) The function
$J_{c}(\Phi/\Phi_{1})$
for the same 1D chains
(using same colors).
The curves for different chains display the same set of peaks,
namely,
at
$\Phi/\Phi_{1}=1$ (first matching field)
and
$\Phi/\Phi_{1}=0.5$,
as well as
at the golden-mean-related values:
$\Phi/\Phi_{1}=\tau$,
$\tau/2$,
$(\tau+1)/2=\tau^{2}/2$,
$(\tau^{2}+\tau)/2=\tau^{3}/2$,
$\tau^{2}=\tau+1$,
$\tau^{2}+1$.
This behavior demonstrates the self-similarity
of
$J_{c}(\Phi)$.
}
\vspace{-0.5cm}
\end{figure}

The results of calculating
$J_{c}(N_{v})$
for chains of different length 
and the same
$\gamma = 1/\tau$
are shown in
Fig.~1a.
The plot clearly shows that,
for sufficiently long chains,
the
{\it positions} of the {\it main peaks in $J_{c}$},
to a significant extent,
do {\it not } depend on the length of the chain.
The peaks
form a Fibonacci sequence:
$N_{v} = 13,$ 21, 34, 55, 89, 144,
and other ``harmonics'':
$N_{v} \cong 17,$ 27.5 ($= 55/2$), 44.5 ($= 89/2$), etc.
Of course,
longer chains allow to better reveal peaks for larger Fibonacci numbers.
In Fig.~1b, the same curves are
rescaled, normalized by the numbers of pins
in each chain.
The rescaled
$J_{c}$
curves
reproduce each other
and have many pronounced peaks for golden-mean-related values
of
$\Phi/\Phi_{1}$
($\Phi_{1}$ is the flux corresponding to the first matching field, $H_{1}$, 
when $N_{v}=N_{p}$),
as shown in Fig.~1b.
Therefore,
the same peaks of
$J_{c}(\Phi)$,
for different chains,
are revealed before and after rescaling
because of the self-similarity of
$J_{c}(\Phi)$.
The self-similarity
of
$J_{c}(\Phi)$
has also been studied in
reciprocal
$k$-space
and will be presented elsewhere \cite{wepenprb}.

{\it Penrose lattice.---}
Consider now a
2D QP APS, namely,
an APS located at the nodes of a
five-fold Penrose lattice.
This lattice is a 2D QP structure, or quasicrystal, 
also referred to as Penrose tiling \cite{bookqc}.
These structures
possess
a perfect
local
rotational (five- or ten-fold) symmetry, 
but
do not have translational long-range order.
The unusual self-similar difraction pattern of a Penrose lattice
exhibits a dense set of ``Bragg'' peaks
because the lattice contains an infinite number of periods in it \cite{bookqc}.
It is precisely this unusual property that is responsible for the striking
$J_{c}(\Phi)$'s obtained here.

\begin{figure}[btp]
\begin{center}
\vspace*{-0.5cm}
\hspace*{-1.5cm}
\includegraphics*[width=11.5cm]{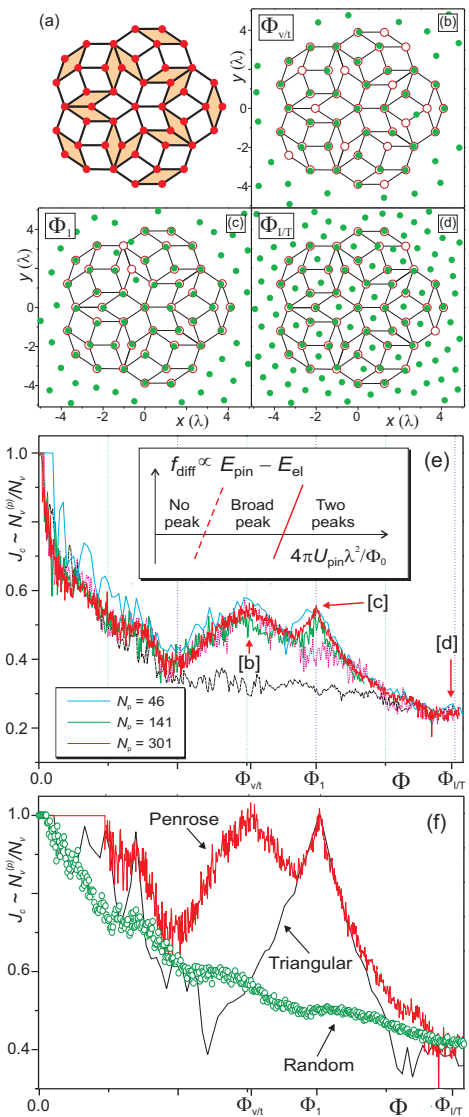}
\end{center}
\vspace{-1.5cm}
\caption{
(Color online)
(a)
An example of a five-fold Penrose lattice,
consisting of
``thick'' and ``thin'' rhombuses.
(b to d)
The
location
of vortices
(green dots)
and
the Penrose-lattice APS
(red open circles
connected by
black
lines)
for:
(b)
$\Phi = \Phi_{\rm vacancy/thin} \equiv \Phi_{\rm v/t} = 0.757 \, \Phi_{1}$,
vortices occupy all the pinning sites
except those in
one of the two vertices of each ``thin'' rhombus;
(c)
$\Phi = \Phi_{1}$,
$N_{v}$
coincides with
$N_{p}$;
(d)
$\Phi = \Phi_{\rm interstitial/thick} \equiv \Phi_{\rm i/T} = 1.482 \, \Phi_{1}$,
vortices occupy both
the pinning sites
and the interstitial positions inside each ``thick'' rhombus.
Here
$f_{p}/f_{0} = 2.0$,
$r_{p} = 0.1\lambda$.
(e)
The
$J_{c}(\Phi \sim N_{v})$
for
Penrose-lattice arrays
with $N_{p} = 46, 141, 301$. 
The data points [b], [c], [d] refer to snapshots in 2b, 2c, 2d, respectively. 
The peak [c] at $\Phi_{1}$ is suppressed for weaker pinning (magenta dotted $J_{c}$ curve).
Eventually, all the main peaks disappear for sufficiently weak pins (black dashed curve). 
The inset shows the dimensionless difference, $f_{\rm diff}$, of the pinning and the elastic 
energies versus the pinning-to-interaction energy ratio, for the broad $J_{c}$ peak 
at $\Phi_{\rm v/t}$ (red dashed line) and for $\Phi_{1}$ (red solid line). 
Only $f_{\rm diff} > 0$ gives stable peaks in $J_{c}$. 
(f)
The
$J_{c}(\Phi)$
for a 301-sites Penrose-lattice (red solid line),
(recalculated for flux only on the Penrose area, $A_{P}$),
triangular (black solid line)
and random (green open circles and solid line)
pinning arrays.
}
\vspace{-0.5cm}
\end{figure}

The structure of a five-fold Penrose lattice is presented in Fig.~2a.
As an illustration only,
a small five-fold symmetric fragment with
$46$
points is shown.
According to specific rules,
the points are connected by lines
in order to display the structure of the Penrose lattice.
The elemental building blocks are
rhombuses
with equal sides
$a$
and angles which are multiples of
$\theta = 36^{\rm o}$.
There are two kinds of rhombuses:
(i) those having angles
$2 \theta$
and
$3 \theta$
(so-called ``thick''; shown empty in Fig.~2a),
and
(ii) rhombuses with angles
$\theta$
and
$4 \theta$
(so-called ``thin''; filled in orange in Fig.~2a).

Let us analyze whether any specific matching effects can exist
between the Penrose pinning lattice and the interacting vortices,
which affect the
magnetic-field dependence of the
critical depinning current $J_{c}(\Phi)$ (Fig.~2). 
Quasicrystalline patterns are intrinsically {\it incommensurate}
with the flux lattice for {\it any} value of the magnetic field
\cite{behrooz},
therefore,
in contrast to periodic (e.g., triangular or square) pinning arrays,
one might
{\it a priori}
assume a {\it lack} of sharp peaks in
$J_{c}(\Phi)$
for QP APS.
However, 
the existence of
many periods in the Penrose lattice
can lead to
a hierarchy of
matching effects for certain values of the applied magnetic field,
resulting in
strikingly-broad shapes for
$J_{c}(\Phi)$.
These could be valuable for applications demanding unusually broad
$J_{c}(\Phi)$'s.

To match the vortex lattice on an entire QP APS,
the specific geometry of the elements which form the QP lattice is important, 
as well as their arrangement.
While the sides, $a$, of the rhombuses are equal,
the distances between the nodes
are
{\it not}
equal.
The lengths of the diagonals of the rhombuses
(Fig.~2a) are:
$1.176a$,
$\tau a \approx 1.618a$
(for the thick rhombus),
$(\tau -1)a=a/\tau \approx 0.618a$,
and $1.902a$ (for the thin rhombus).
Based on these distances,
we can predict matching effects
(and corresponding features of the function
$J_{c}(\Phi)$)
for the Penrose-lattice APS.

First,
there is a ``first matching field''
(we denote the corresponding flux as $\Phi_{1}$)
when each pinning site is occupied by a vortex (Fig.~2c).
Although the sides of all the rhombuses are equal
to each other, 
nevertheless this matching effect is not expected to be
accompanied by a sharp peak.
Instead, it is a broad maximum (peak [c] in Fig.~2e)
involving three kinds of local
``commensurability'' effects of the flux lattice:
with the rhombus side $a$;
with the short diagonal of a thick rhombus,
$1.176a$, which is close to $a$;
and with the short diagonal of a thin rhombus,
which is $a/\tau \approx 0.618a$.
For the 
overall 
square cell used,
some of the vortices are outside the Penrose sample;
these mimic the applied magnetic field and determine
the average vortex density in the entire cell.
Because of the additional vortices outside the Penrose APS, 
our computed 
$J_{c}(\Phi)$
is
reduced
by a factor
$\eta = A_{P}/A \approx 0.575$ (see Fig.~2e),
where $A_{P}$ and $A$ are the areas
of the Penrose lattice and of the cell.

Another matching (Fig.~2b) is related with 
the filling of all the pinning sites on the vertices of the thick rhombuses
and only
{\it three out of four}
of the pinning sites on the vertices of thin rhombuses,
i.e.,
one of the pinning sites
on the vertices of the thin rhombuses
is empty.
For this value of the flux,
matching conditions are fulfilled for
{\it two }
close distances,
$a$ (the side of a rhombus)
and
$1.176a$ (the short diagonal of a thick rhombus), 
but
are not fulfilled for
the short diagonal, $a/\tau$, of the thin rhombus.
Therefore,
this
2D QP feature
is related to
$\tau$,
although not in such a
direct
way as in the case of a 1D QP pinning array.
This 2D QP matching results in a very wide maximum (arrow [b] in Fig.~2e) of the function
$J_{c}(\Phi)$.
The position of this broad maximum
(denoted here by $\Phi_{\rm vacancy/thin} \equiv \Phi_{\rm v/t} = 0.757 \, \Phi_{1}$)
could be found as follows.
The ratio of the numbers of thick and thin rhombuses is determined by the 
Fibonacci numbers and in the limit of large pinning arrays, 
$N_{p} \to \infty$, this ratio tends to $\tau$.
The number of unoccupied pinning sites is governed by the number of thin rhombuses.
However, some of the thin rhombuses are separated from other thin rhombuses by a single thick one
({\it single}
thin rhombuses; see Fig.~2a),
while some of thin rhombuses have common sides with each other
(orange arrow-shaped {\it double} thin rhombuses in Fig.~2a).
Therefore, the number of vacancies (i.e., unoccupied pins) is then the number of single thin rhombuses
$N_{\rm rh}^{\rm s}$ plus one half
of the number $N_{\rm rh}^{\rm d}$ of ``double'' thin rhombuses,
$
N_{p}^{\rm un} (\Phi_{\rm v/t}) = N_{\rm rh}^{\rm s} + N_{\rm rh}^{\rm d}/2,
$
where
$N_{p}^{\rm un}$
is the number of unoccupied pinning sites at
$\Phi = \Phi_{\rm v/t}$.

For higher vortex densities
($\Phi = \Phi_{\rm interstitial/thick} \equiv \Phi_{\rm i/T} = 1.482 \, \Phi_{1}$)  
a single interstitial vortex 
is inside 
each thick rhombus (see Fig.~2d). 
These interstitial vortices can easily move; thus $J_{c}$ has no peak at $\Phi_{\rm i/T}$. 
The position of this feature is determined by the number of vortices at 
$\Phi = \Phi_{1}$,
which is
$N_{v}(\Phi) = N_{p}$,
plus
the number of thick rhombuses,
$N_{\rm rh}^{\rm thick} = N_{\rm rh}/\tau$.


In order to better understand the 
structure of $J_c(\Phi)$ for the Penrose pinning 
lattice, we compare the elastic $E_{\rm el}$ and 
pinning $E_{\rm pin}$ energies of the vortex lattice at 
$H_1$ and at (the lower field) $H_{\rm v/t}$, 
corresponding to the two maxima of $J_c$ (Figs.~2e,f). 
Vortices can be pinned 
if the gain $E_{\rm pin}= U_{\rm pin} \, \beta \, n_{\rm pin}$
of the pinning energy is larger than the increase of the elastic
energy 
\cite{brandt}
related to local compressions: 
$E_{\rm el}=C_{11}[(a_{\rm eq}-b)/a_{\rm eq}]^2$. 
The shear elastic energy ($\propto C_{66}$) provides the same qualitative result \cite{wepenprb}. 
Here, $U_{\rm pin}\sim f_p \, r_p$, 
$n_{\rm pin}$ is the density of pinning centers, 
$\beta (H \leq H_{1}) = H/H_{1} = B/(\Phi_{0} n_{\rm pin})$, and $\beta (H > H_{1}) = 1$ 
is the fraction of occupied pinning sites
($\beta=1$ for $H=H_1$, and $\beta=0.757$ for $H=H_{\rm v/t}$), 
$a_{\rm eq}=\left( 2/\sqrt{3}\beta n_{\rm pin} \right)^{1/2}$ is the equilibrium 
distance between vortices in the triangular lattice, 
$b$ is the minimum distance between
vortices in the distorted pinned vortex lattice ($b=a/\tau$ for
$H=H_1$ and $b=a$ for $H=H_{\rm v/t}$), and
$C_{11}=B^2/[4\pi(1+\lambda^2 k^2)]$ is the compressibility modulus
for short-range deformations 
\cite{brandt}
with characteristic spatial scale 
$k \approx \left( n_{\rm pin} \right)^{1/2}$. 
The dimensionless difference of
the pinning and elastic energies is 
$E_{\rm pin}-E_{\rm el}=\beta f_{\rm diff} \, n_{\rm pin} \Phi_0^2/(4\pi\lambda^2)$, 
where
$
f_{\rm diff}=4\pi\lambda^2 U_{\rm pin} / \Phi_0^2 
- \beta \left[ 1 - b \left( \beta \sqrt{3} n_{\rm pin} / 2 \right)^{1/2} \right]^2.
$
Near matching fields, $J_c$ has a peak when $f_{\rm diff} > 0$
(and no peak when $f_{\rm diff} < 0$). 
Since only two matching fields provide $f_{\rm diff} > 0$, 
then our analysis explains the two-peak structure observed in $J_c$ 
shown in Figs.~2e,f. 
For instance, for the main matching fields:
$f_{\rm diff}(\Phi_{\rm v/t}) \approx 0.0056$, 
$f_{\rm diff}(\Phi_{1}) \approx 0.0058$, 
and 
$f_{\rm diff}(\Phi_{\rm i/T}) \approx -0.09$. 
Note that for weaker pinning, the two-peak structure gradually turns into 
one very broad peak, and eventually zero peaks  for weak enough pinning (see Fig.~2e). 
The $J_{c}$ peaks corresponding to higher
matching fields are strongly suppressed because of the fast increase
($\propto B^2$) of the compressibility modulus $C_{11}$ and, thus, the elastic
energy with respect to the pinning energy; 
the latter cannot exceed the maximum value $U_{\rm pin}n_{\rm pin}$. 
The subharmonic peaks of
$J_c$, which could occur for lower fields $H<H_{\rm v/t}$, are also
suppressed due to the increase of $C_{11}$ associated with the growing
spatial scales $1/k$ of the deformations. 


For comparison, we show the
$J_{c}(\Phi)$ for the Penrose-lattice (itself, i.e. calculated only for $A_{P}$),
triangular and random pinning arrays (Fig.~2f). 
The latter is an average over five realizations of disorder. 
Notice that the QP
lattice leads to a very broad and potentially useful {\it
enhancement} of the critical current $J_{c}(\Phi)$, even compared
to the triangular or random APS. The remarkably broad maximum in
$J_{c}(\Phi)$ is due to the fact that the Penrose lattice has {\it
many} (infinite, in the thermodynamic limit) periodicities built
in it \cite{bookqc}. 
In principle, 
each one of these periods provides a peak in $J_{c}(\Phi)$. 
In practice, 
like in 
quasicrystalline 
difraction patterns, only few peaks are strong. 
This is also consistent with our study. 
Furthermore, 
the pinning parameters can be 
adjusted by using as pinning centers either antidots 
``drilled'' in the film 
\cite{vvmdotprl,rwdot}, or blind antidots \cite{vvmbdot} of
different depths and radii. Thus, our results could be 
observed experimentally.


{\it Conclusions.---} The critical depinning current $J_{c}(\Phi)$
was studied in 1D QP chains and in 2D QP arrays (the five-fold
Penrose lattice) of pinning sites. A hierarchical and self-similar
$J_{c}(\Phi)$ was obtained. We 
physically analyzed 
all the main features of
$J_{c}(\Phi)$. Our analysis shows that the QP lattice provides an
unusually broad critical current $J_{c}(\Phi)$, that could be
useful for practical applications demanding high $J_{c}$'s over a
wide range of fields.
Our proposal can be easily extended, mutatis mutandi, to other
related systems, including colloidal suspensions interacting with
pinning traps provided by arrays of optical tweezers~\cite{Grier}.

\smallskip

This work was supported in part by 
ARDA 
and NSA under AFOSR contract 
F49620-02-1-0334; and by the US 
NSF 
grant No.~EIA-0130383.

\end{document}